\documentclass[twocolumn]{aastex631}

\def\specialname[#1]{\textbf{\textsc{#1}}}

\usepackage{multirow}
\usepackage{soul}

\shorttitle{The stellar mass-halo mass relation}
\shortauthors{Wang et al.}

\graphicspath{{./}{figures/}}

\begin{document}

\title{Testing galaxy formation models with the stellar mass-halo mass relations for star-forming and quiescent galaxies}

\author[0000-0002-3775-0484]{Kai Wang}\thanks{KW: wkcosmology@gmail.com}
\affiliation{Kavli Institute for Astronomy and Astrophysics, Peking University, Beijing 100871, China}
\affiliation{Institute for Computational Cosmology, Department of Physics, Durham University, South Road, Durham, DH1 3LE, UK}
\affiliation{Centre for Extragalactic Astronomy, Department of Physics, Durham University, South Road, Durham DH1 3LE, UK}

\author{Yingjie Peng}\thanks{YP: yjpeng@pku.edu.cn}
\affiliation{Kavli Institute for Astronomy and Astrophysics, Peking University, Beijing 100871, China}
\affiliation{Department of Astronomy, School of Physics, Peking University, Beijing 100871, China}

\begin{abstract}
    The tight relationship between the stellar mass and halo mass of galaxies
    is one of the most fundamental scaling relations in galaxy formation and
    evolution. It has become a critical constraint for galaxy formation models.
    Over the past decade, growing evidence has convincingly shown that the
    stellar mass-halo mass relations (SHMRs) for star-forming and quiescent
    central galaxies differ significantly: at a given stellar mass, the average
    host halo mass of quiescent centrals is more massive than that of the
    star-forming centrals. Despite the importance of this feature, its
    scientific implications have not yet been fully recognized or thoroughly
    explored in the field. In this work, we demonstrate that the
    semi-analytical model L-GALAXIES successfully reproduces these
    observational results, whereas three state-of-the-art hydrodynamic galaxy
    formation simulations (TNG, Illustris, and EAGLE) do not. Consequently, in
    L-GALAXIES, star-forming central galaxies are more efficient at converting
    baryons into stars than quiescent central galaxies at a given halo mass,
    while the other models predict similar efficiencies for both populations.
    Further analysis reveals that these fundamental discrepancies stem from
    distinct evolutionary paths on the stellar mass-halo mass plane. We show
    that the observed SHMRs for star-forming and quiescent galaxies support
    galaxy formation models in which quenching only weakly correlates with halo
    assembly histories, and in which the stellar mass of star-forming galaxies
    can increase significantly since cosmic noon. In contrast, models in which
    quenching strongly prefers to happen in early-formed halos are not very
    favored. Additionally, we find that galaxy downsizing is present in
    L-GALAXIES and TNG, but absent in Illustris and EAGLE.
\end{abstract}

\keywords{method: statistical - galaxies: evolution - galaxies: formation - galaxies: halos - galaxies: groups: general}

\section{Introduction}%
\label{sec:introduction}

Galaxies are formed and evolved within dark matter halos, where baryons are
accreted by dark matter halos to fuel star formation, and the converted stellar
mass is limited by the total available baryons, which scales with halo mass.
Regarding the intimate interplay between star formation and halo growth, a
tight scaling relation between the stellar mass of galaxies and the host halo
mass, named as the stellar mass-halo mass relation, is expected
\citep{coorayHaloModelsLarge2002, baughPrimerHierarchicalGalaxy2006,
    shankarNewRelationshipsGalaxy2006, moGalaxyFormationEvolution2010,
wechslerConnectionGalaxiesTheir2018}. In fact, numerous efforts in the last
decades have been devoted to measuring SHMR in observation, based on various
techniques, like gravitational lensing
\citep{mandelbaumStrongBimodalityHost2016, bilickiBrightGalaxySample2021,
zhangMassiveStarformingGalaxies2022}, satellite kinematics
\citep{moreSatelliteKinematicsIII2011, liInternalKinematicsGroups2012,
liStellarMassStellar2013, zhangMassiveStarformingGalaxies2022}, dynamical
modelling \citep{lapiPrecisionScalingRelations2018, postiPeakStarFormation2019,
    romeoLenticularsBlueCompact2020,
    postiDynamicalEvidenceMorphologydependent2021,
diteodoroDarkMatterHalos2023}, etc. The most remarkable finding is that all
different methods deliver highly convergent scaling relation between
stellar mass and halo mass, despite different assumptions on which they are
based.

The observed SHMR tells us that the stellar mass to halo mass ratio, which is
the efficiency of converting baryons into stars, is low overall, which means
that the star formation process is quite inefficient. In addition, the
efficiency is suppressed at the low- and high-mass end, which suggests at least
two mechanisms to suppress the star formation efficiencies, operating in
low-mass and high-mass halos separately
\citep{yangConstrainingGalaxyFormation2003, yangEvolutionGalaxyDarkMatter2012,
    conroyCONNECTINGGALAXIESHALOS2009,
    behrooziCOMPREHENSIVEANALYSISUNCERTAINTIES2010,
mosterGalacticStarFormation2013}. This result provokes numerous observational
and theoretical studies on the feedback effect from stellar wind, supernovae
explosion, and the accretion of super-massive black holes in galaxy center, and
their impact on the global galaxy properties, such as the star formation rate
\citep{kauffmannFormationEvolutionGalaxies1993,
kangSemianalyticalModelGalaxy2005, crotonManyLivesActive2006}. It also leads to
the proposal of other mechanisms to suppress stellar conversion efficiency,
which is exemplified by the inefficient gas cooling in massive dark matter
halos \citep{whiteCoreCondensationHeavy1978,
whiteGalaxyFormationHierarchical1991, dekelColdStreamsEarly2009}.

Regarding the impactful findings elicited by measuring and understanding the
SHMR for the general population of galaxies, several studies have initiated the
measurement of SHMRs for different populations of galaxies, such as
star-forming galaxies versus quiescent galaxies, and they converged to the same
conclusion: at fixed stellar mass, quiescent/red central galaxies on average
occupy more massive dark matter halos than star-forming/blue central galaxies.
In other words, {\it at fixed stellar mass, star-forming/blue central galaxies
are more efficient in converting their available baryons into stars}
\citep[e.g.][]{moreSatelliteKinematicsIII2011,
    mandelbaumStrongBimodalityHost2016, zuMappingStellarContent2016,
    postiPeakStarFormation2019, postiDynamicalEvidenceMorphologydependent2021,
zhangMassiveStarformingGalaxies2022, kakosStarformingQuiescentCentral2024a}.
However, the scientific implications of these observational results have been
inadequately discussed in literature. In this work, we start with the
comparison between observational results and galaxy formation models, pointing
out that most of the state-of-the-art galaxy formation models fail to reproduce
the observational result, and discuss its scientific implications on the
relationship between star formation histories and halo growth histories, as
well as stringent constraints that can be put on galaxy formation models.

This paper is structured as follows: \S\,\ref{sec:data} introduces
observational results on the SHMR of star-forming and quiescent galaxies and
galaxy formation models. \S\,\ref{sec:result} presents the comparison between
observations and models. \S\,\ref{sec:discussion} discusses the implication on
the modeling of galaxy formation and evolution, followed by a summary in
\S\,\ref{sec:summary}. Throughout this paper, we adopt a concordance cosmology
with $\Omega_\Lambda = 0.7$, $\Omega_{\rm m} = 0.3$, and $h = 0.7$. All
logarithms are 10-based.

\section{Data}%
\label{sec:data}

\subsection{Observations}%
\label{sub:observations}

We compile observational measurements of the SHMRs for different populations of
galaxies in literature and present them on the first two panels in
Fig.~\ref{fig:paper_compare} with a brief introduction presented as follows.

\paragraph{Satellite kinematics}%
\label{par:satellite_kinematics}

\citet{moreSatelliteKinematicsIII2011} measure the halo mass of blue and red
central galaxies selected from SDSS survey using the kinematics of surround
satellite galaxies \citep[see also][]{zhangMassiveStarformingGalaxies2022}.

\paragraph{Weak gravitational lensing}%
\label{par:weak_gravitational_lensing}

\citet{mandelbaumStrongBimodalityHost2016} measure the halo mass of blue and
red local brightest galaxies (LBG)\footnote{Local brightest galaxies are
    selected as those galaxies with no brighter neighbors within 1 Mpc and
$\pm$ 1000km/s.}, separately, using the weak gravitational lensing
technique in SDSS \citep[see also][]{bilickiBrightGalaxySample2021,
zhangMassiveStarformingGalaxies2022}.

\paragraph{Rotation curve}%
\label{par:rotation_curve}

\citet{lapiPrecisionScalingRelations2018} compile a sample of 550
disk-dominated galaxies with rotation curve measurements, from which they can
estimate their host halo mass. \citet{postiPeakStarFormation2019} measure the
halo mass of 108 spiral galaxies with HI rotation curve measurements.
\citet{diteodoroDarkMatterHalos2023} perform the 3D kinematic modeling for 15
massive spiral galaxies in the local Universe.

\paragraph{Globular cluster systems}%
\label{par:globular_cluster_systems}

\citet{postiDynamicalEvidenceMorphologydependent2021} measure the halo mass of
25 early-type galaxies based on the kinematics of their globular cluster
systems.

\subsection{Galaxy formation models}%
\label{sub:galaxy_formation_models}

\paragraph{Semi-analytical modeling}%
\label{par:semi_analytical_modeling}

\texttt{L-GALAXIES} is a semi-analytical model of galaxy formation, built on
halo merger trees extracted from the Millennium N-body cosmological
simulations. Galaxies are seeded within dark matter halos, and then the accreted
baryons, along with the growth of dark matter halo, are used to feed star
formation, once the accreted gas is sufficiently cooled down and compressed. It
also models various astrophysical processes that affect global galaxy
properties, such as stellar feedback, metal production, the growth and feedback
from the super-massive black hole located at the center of each galaxy, the
merging of galaxies and subsequent starbursts, and others. The free parameters
in the model recipes are constrained by the stellar mass function and the
fraction of passive galaxies across the redshift range from 0 to 3. We refer to
\citet{henriquesGalaxyFormationPlanck2015} for more details \citep[see
also][]{guoDwarfSpheroidalsCD2011}.

\paragraph{Hydrodynamical simulations}%
\label{par:hydrodynamical_simulations}

\texttt{IllustrisTNG} \citep[hereafter
\texttt{TNG};][]{weinbergerSimulatingGalaxyFormation2017,
    naimanFirstResultsIllustrisTNG2018, marinacciFirstResultsIllustrisTNG2018,
    nelsonFirstResultsIllustrisTNG2018, pillepichSimulatingGalaxyFormation2018,
    pillepichFirstResultsIllustrisTNG2018,
nelsonIllustrisTNGSimulationsPublic2019}, \texttt{Illustris}
\citep{vogelsbergerPropertiesGalaxiesReproduced2014,
    vogelsbergerIntroducingIllustrisProject2014,
    genelIntroducingIllustrisProject2014,
    sijackiIllustrisSimulationEvolving2015,
nelsonIllustrisSimulationPublic2015}, and \texttt{EAGLE}
\citep{schayeEAGLEProjectSimulating2015,
mcalpineEagleSimulationsGalaxy2016, theeagleteamEAGLESimulationsGalaxy2017}
are three state-of-the-art cosmological hydrodynamical galaxy formation
simulations, in which dark matter, stars, gas, and super-massive black hole
are represented by discrete particles, whose evolution should be governed
by fundamental physical laws. Nevertheless, given the limited computational
power, these simulations still need empirical recipes to model physical
processes that are below the resolution limit but have an impact on larger
scales. These processes are exemplified by star formation, stellar
feedback, and the growth and feedback of super-massive black holes, which
are also the major factors that differentiate different simulations. We
refer to specific papers on these simulations for more details.

For all simulations, dark matter halos are identified using the
Friends-of-Friends (FoF) algorithm
\citep{davisEvolutionLargescaleStructure1985}, and subhalos are identified
using the \texttt{SUBFIND} algorithm
\citep{springelPopulatingClusterGalaxies2001} in each FoF halo. Across redshift
snapshots, merger trees are constructed using the algorithm in
\citet{springelSimulationsFormationEvolution2005}, and the main branch is
constructed by recursively selecting the progenitor subhalo with the most
massive history \citep{deluciaHierarchicalFormationBrightest2007}. The halo
mass is defined as the total mass enclosed in a radius within which the mean
density just exceeds 200 times the critical density at the given redshift. The
stellar mass is defined as the recommended way in each simulation, which is the
total stellar mass within 30 physical kpcs in \texttt{EAGLE} and within twice
the half-mass radius in \texttt{Illustris} and \texttt{TNG}\footnote{They first
    found all stellar particles that are gravitational bound to each subhalo,
    from which they could measure the half-mass radius. Then they measure the
total stellar mass as those nclosed in twice the half-mass radius.}. The
same aperture is also applied to the calculation of star formation rate
(SFR) in each simulation.

\subsection{Separating two populations of galaxies}%
\label{sub:separating_two_populations_of_galaxies}

It is noteworthy that different studies separate galaxies into two populations
based on different types of galaxy bi-modality. For instance,
\citet{mandelbaumStrongBimodalityHost2016} discriminate blue and red galaxies
based on their $g-r$ color; so do \citet{moreSatelliteKinematicsIII2011}, while
\citet{bilickiBrightGalaxySample2021} uses the $u-g$ color.
\citet{lapiPrecisionScalingRelations2018} and
\citet{postiDynamicalEvidenceMorphologydependent2021} separate early-type
galaxies from late-type galaxies based on their morphology \citep[see
also][]{postiPeakStarFormation2019}.  In this paper, we define star forming
galaxies as those with ${\rm SFR}/M_* > 10^{-11}{\rm  yr^{-1}}$ and quiescent
galaxies as the remaining ones, in all four galaxy formation models. 

\section{Result}%
\label{sec:result}

\begin{figure*}
    \centering \includegraphics[width=\linewidth]{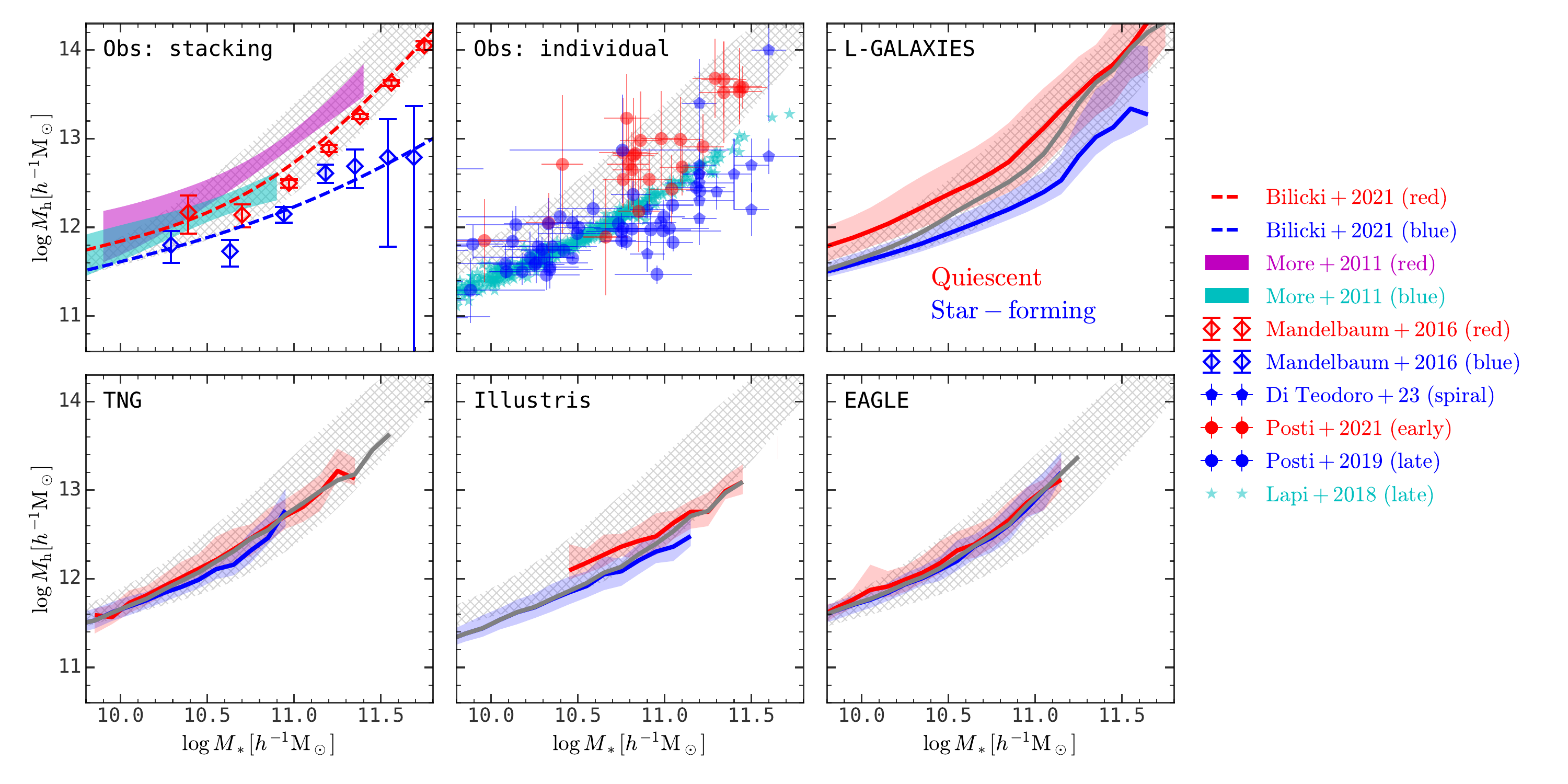}
    \caption{
        The stellar mass-halo mass relation for star-forming and quiescent
        central galaxies. The top-left panel is for observational results from
        weak gravitational lensing
        \citep[see][]{mandelbaumStrongBimodalityHost2016,
        bilickiBrightGalaxySample2021} and satellite kinematics
        \citep[see][]{moreSatelliteKinematicsIII2011}. The top-middle panel is
        for the measurement of individual galaxies, where the halo mass is
        constrained by rotation curves
        \citep{lapiPrecisionScalingRelations2018, postiPeakStarFormation2019,
        diteodoroDarkMatterHalos2023} and globular cluster systems
        \citep{postiDynamicalEvidenceMorphologydependent2021}. The remaining
        four panels are for the results of four different galaxy formation
        models: \texttt{L-GALAXIES}, \texttt{TNG}, \texttt{Illustris}, and
        \texttt{EAGLE}, with red/blue/gray solid lines show the median halo
        mass of quiescent/star-forming/total galaxies as a function of stellar
        mass. The hatched region in all panels shows the $16^{\rm th}-84^{\rm
        th}$ quantiles of halo mass as a function of stellar mass for all
        central galaxies in \texttt{L-GALAXIES} for reference. This figure
        demonstrates that star-forming and quiescent central galaxies follow
        different stellar mass-halo mass scaling relations, and
        \texttt{L-GALAXIES} can reproduce this trend while the other three
        hydrodynamical galaxy formation simulations fail to do so.
    }%
    \label{fig:paper_compare}
\end{figure*}

\begin{figure*}
    \centering
    \includegraphics[width=1\linewidth]{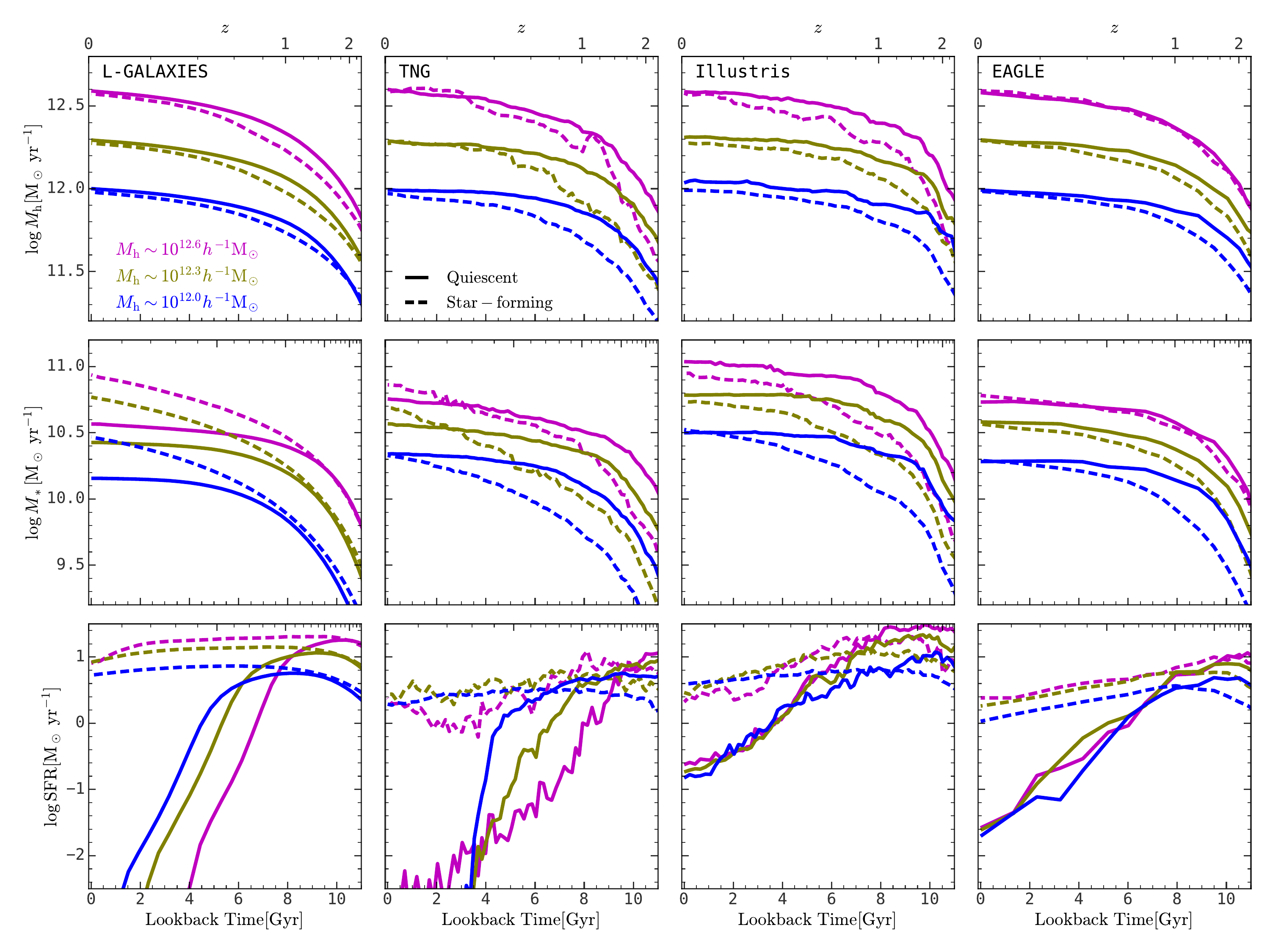}
    \caption{
        The redshift evolution of halo mass (top panels), progenitor stellar
        mass (middle panels), and star formation rate (bottom panels) for
        quiescent (solid) and star-forming (dashed) central galaxies selected
        at $z=0$ in four different galaxy formation models. Here central
        galaxies are binned into three $z=0$ halo mass bins and each bin has a
        width of 0.2 dex.
    }%
    \label{fig:paper_evolution_history}
\end{figure*}

\begin{figure*}
    \centering
    \includegraphics[width=1\linewidth]{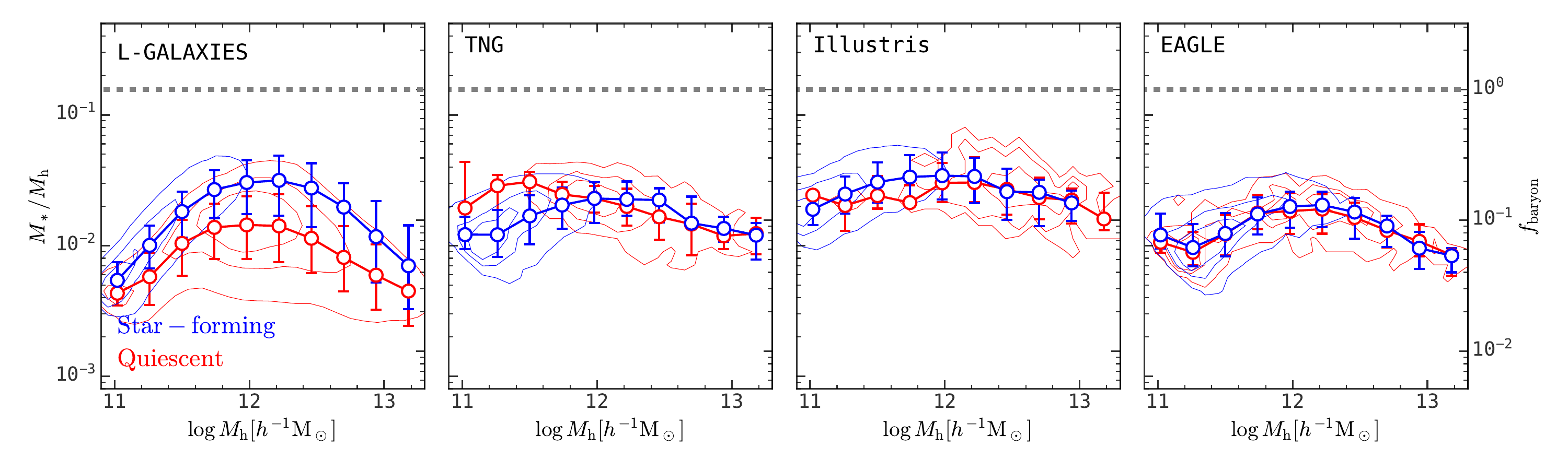}
    \caption{
        The central stellar mass-to-halo mass ratio as a function of halo mass
        for four different galaxy formation models at $z=0$. The background
        contour lines enclose 30\%, 60\%, and 90\% star-forming and quiescent
        central galaxies in blue and red, respectively. The error bars show the
        median and $16^{\rm th}-84^{\rm th}$ quantiles of the ratio in each
        halo mass bin. The y-axis on the right side shows $f_{\rm
        baryon}=\Omega_{\rm m}/\Omega_{\rm b}\times M_*/M_{\rm h}$. This figure
        shows that, in \texttt{L-GALAXIES}, star-forming central galaxies are
        more efficient in converting baryons into stars than quiescent central
        galaxies, while other models give similar efficiencies for these two
        populations.
    }%
    \label{fig:paper_conversion_eff}
\end{figure*}

\begin{figure*}
    \centering
    \includegraphics[width=1\linewidth]{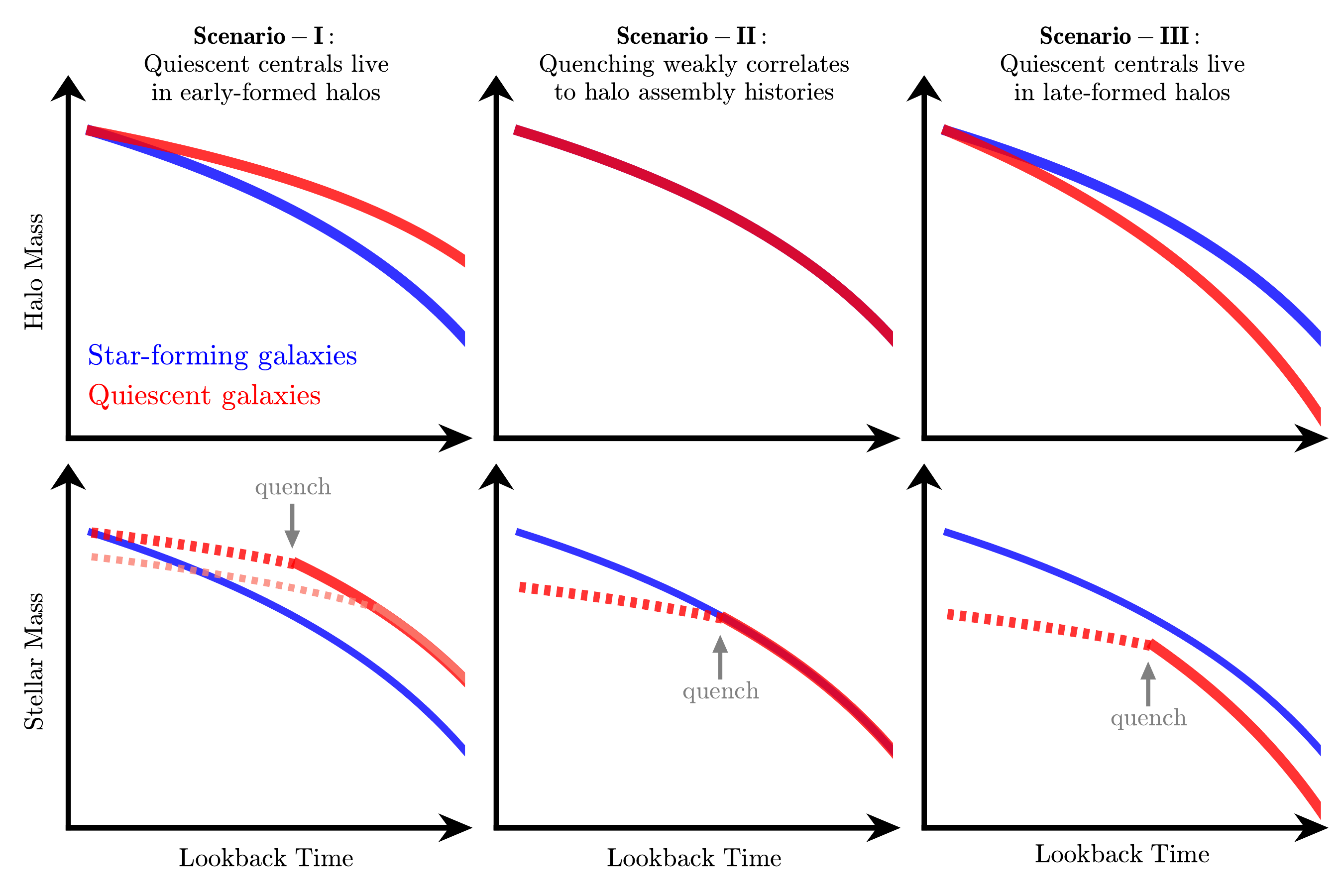}
    \caption{
        Illustration of the redshift evolution of halo mass and stellar mass
        for star-forming (blue) and quiescent (red) descendant galaxies at
        $z=0$. In \textbf{Scenario-I} (left panels), quiescent central galaxies
        at $z=0$ prefer to live in early-formed halos. In this case, the
        progenitor galaxies of quiescent central galaxies were initially more
        massive since their progenitor halos are more massive and most of the
        central galaxies at high redshifts are star-forming. However, once
        these galaxies are quenched, their stellar growth will be suppressed
        and, consequently, these two populations of galaxies end up with
        similar stellar masses. In \textbf{Scenario-II} (middle panels),
        quiescent central galaxies live in halos with similar assembly
        histories as star-forming central galaxies. In this case, the
        progenitor galaxies of these two populations of galaxies have similar
        stellar mass at early times. Then, once some galaxies are quenched,
        their stellar mass growth will be suppressed and end up with lower
        stellar mass compared with their star-forming counterparts, which trend
        is consistent with observational results. In \textbf{Scenario-III}
        (right panels), quiescent central galaxies live in late-formed halos.
        In this case, the progenitor galaxies of quiescent central galaxies not
        only possess lower stellar mass at early times, but also are suppressed
        in their stellar mass growth after quenching, compared with their
        star-forming counterparts. Consequently, the stellar mass difference
        becomes larger.
    }%
    \label{fig:figures/paper_illustration2}
\end{figure*}

\subsection{Comparing models to observations}%
\label{sub:comaring_models_to_observations}

There have been numerous efforts devoted to measuring the SHMRs of star-forming
and quiescent central galaxies in our local Universe, and a compilation of
these observational results is shown in Fig~\ref{fig:paper_compare}. The
top-left panel shows the measurements based on stacking systems with similar
stellar mass together to measure their halo properties, which is exemplified by
weak gravitational lensing \citep[e.g.][]{mandelbaumStrongBimodalityHost2016,
bilickiBrightGalaxySample2021, zhangMassiveStarformingGalaxies2022} and
satellite kinematics \citep[e.g.][]{moreSatelliteKinematicsIII2011,
langeUpdatedResultsGalaxyhalo2019, zhangMassiveStarformingGalaxies2022}. All
these results, despite their inherent systematics, agree on the qualitative
conclusion that red and quiescent central galaxies tend to live in more massive
dark matter halos than their blue and star-forming counterparts at fixed
stellar mass. The same trend can also be inferred from the top-middle panel,
where the halo mass is measured for each individual galaxy by modeling either
the rotation curves or the globular cluster systems
\citep[e.g.][]{lapiPrecisionScalingRelations2018, postiPeakStarFormation2019,
postiDynamicalEvidenceMorphologydependent2021, diteodoroDarkMatterHalos2023}.
Moreover, we can see that those late-type galaxies follow a single power-law
relation on the stellar mass-halo mass plane \citep[see
also][]{postiPeakStarFormation2019}, and this differs from the SHMR for all
central galaxies, whose slope steepens at the massive end.

The remaining four panels show the results from four state-of-the-art galaxy
formation models, including one semi-analytical model (\texttt{L-GALAXIES}) and
three cosmological hydrodynamical galaxy formation simulations (\texttt{TNG},
\texttt{Illustris}, and \texttt{EAGLE}). To begin with, it should be emphasized
that current cosmological hydrodynamical galaxy formation simulations are not
capable of delivering physical stellar mass due to the insufficient resolution
and the poor knowledge of the feedback process, thus calibration against
observational stellar mass function is required. Therefore, it is meaningless
to inspect the difference in their SMHRs, which only reflects the efforts they
put into calibrating the model, and indeed there are some discrepancies between
different models in Fig.~\ref{fig:paper_compare}. However, the SHMRs for
different populations of galaxies are less manipulated during the calibration
process, hence it is a better test-bed for the physical recipes adopted in the
simulation. As one can see here, at fixed stellar mass, \texttt{L-GALAXIES}
produces a large dispersion of halo mass, while the SHMRs in all three
hydrodynamical simulations are very tight, leaving no room for a difference in
halo mass for star-forming and quiescent central galaxies. Finally,
\texttt{L-GALAXIES} predicts that quiescent central galaxies live in more
massive halos than star-forming central galaxies at fixed stellar mass, in
consistency with observational results \citep[see
also][]{wangSatelliteAbundancesBright2012, mandelbaumStrongBimodalityHost2016},
while all three hydrodynamical simulations predict indistinguishable SHMRs for
these two populations \citep[see][]{marascoMassiveDiscGalaxies2020,
zhangMassiveStarformingGalaxies2022}, except for a minor difference in
\texttt{Illustris}.

\subsection{Relating star formation history to halo assembly history}%
\label{sub:relating_star_formation_history_to_halo_assembly_history}

We have seen that, among four galaxy formation models we inspected, only
L-GALAXIES can reproduce the difference in the stellar mass-halo mass relation
for star-forming and quiescent galaxies. We then proceed to comprehend this
phenomenon by tracking the stellar mass and halo mass growth histories in all
four models at fixed halo mass at $z=0$ for star-forming and quiescent central
galaxies separately. At first, we want to point out that the observational
results indicate that, at fixed descendant halo mass, quiescent central
galaxies have lower stellar mass than their star-forming counterparts,
especially from those galaxies with individual halo mass measurements (see the
top middle panel in Fig.~\ref{fig:paper_compare}). Similar conclusions are
obtained in previous studies based on galaxy clustering analysis \citep[see][]{
rodriguez-pueblaStellartoHaloMassRelation2015}, extended halo occupation
modelling \citep[see][]{xuConditionalColourMagnitude2018} and forward satellite
kinematics modelling \citep[see][]{moreSatelliteKinematicsIII2011}. We note
that this conclusion cannot be made solely from observationally inferred
average halo mass at fixed stellar mass, as demonstrated in previous studies
\citep[e.g.][]{zuMappingStellarContent2016,
    behrooziUNIVERSEMACHINECorrelationGalaxy2019,
mosterEMERGEEmpiricalConstraints2020, tinkerSelfCalibratingHaloBasedGroup2021}.
This is because that, even if the average halo mass as a function of stellar
mass is fixed, the average stellar mass as a function of halo mass is still
dependent on the joint distribution of stellar mass and halo mass, which is
poorly understood and constrained. One way to make this inference is through
individual halo mass measurement and the available measurements are those
systems shown on the top-middle panel of Fig.~\ref{fig:paper_compare}. 

Fig.~\ref{fig:paper_evolution_history} shows the evolution of halo mass,
stellar mass, and star formation rate along the main branch of the galaxy
merger tree as a function of lookback time for star-forming and quiescent
central galaxies at fixed descendant halo mass. We start from the halo growth
history, where all galaxy formation models predict that quiescent central
galaxies prefer to live in early-formed halos. However, the difference in the
halo growth history for star-forming and quiescent central galaxies is
significant in three hydrodynamical simulations, but marginal in
\texttt{L-GALAXIES}. In other words, all three hydrodynamical simulations
predict strong correlations between the halo growth history and the descendant
star formation status, while \texttt{L-GALAXIES} gives a weak correlation.

We then move on to the stellar mass growth history, as shown in the middle
panels in Fig.~\ref{fig:paper_evolution_history}, and we find significant
differences among different galaxy formation models. In \texttt{L-GALAXIES},
star-forming and quiescent galaxies have almost identical stellar mass history
initially, when their progenitor galaxies were both star-forming. Then, for
whatever reason\footnote{In \texttt{L-GALAXIES}, massive central galaxies are
    quenched by the feedback from accreting super-massive black holes in the galaxy
    center, since it will heat the cold gas content, making it unavailable
for subsequent star formation.}, some galaxies are quenched, and their
stellar mass history flattened, while the remaining galaxies were still
actively making new stars to increase their stellar mass. Consequently,
when evolved to $z=0$, star-forming galaxies end up with stellar mass
$0.3-0.4$ dex higher than those quiescent galaxies on average. This result
is consistent with the picture proposed in
\citet{pengMASSENVIRONMENTDRIVERS2012} to explain the halo mass difference
for red and blue central galaxies in observation \citep[see
also][]{manFundamentalRelationHalo2019}.

For \texttt{TNG} and its precedence, \texttt{Illustris}, as shown on the panels
in the middle two columns, the halos that host quiescent descendant central
galaxies are assembled much earlier than those with star-forming central
galaxies. Then, the progenitor stellar mass of quiescent central galaxies grows
much faster than that of star-forming central galaxies at the early
stage, benefiting from their more massive progenitor halos and more available
baryons at high-$z$. Then, when these galaxies are quenched, their stellar mass
growth is suppressed and the advantage in stellar mass accumulated at early
times will be gradually consumed. Eventually, the stellar mass of quiescent
galaxies become very close to that of star-forming galaxies, or could even be
surpassed by the latter, as one can see in \texttt{TNG}. For \texttt{EAGLE},
galaxies and their dark matter halos evolve similarly.

A similar conclusion could be obtained from reading the evolution of SFR for
star-forming and quiescent galaxies, as shown on the bottom panels of
Fig.~\ref{fig:paper_evolution_history}. In \texttt{L-GALAXIES}, the SFR
histories\footnote{This SFR history is different from the star formation
    history we usually refer to in spectral energy distribution fitting, in
    which case all progenitor branches are considered, and here we only include
the main branch.} were identical for these two populations, then, after a
while, the progenitor SFR of quiescent galaxies damped, while the remaining
galaxies are still actively forming new stars. For all three hydrodynamical
galaxy formation simulations, the progenitors of quiescent galaxies have
much higher SFR than those of star-forming galaxies at the initial stage,
and the difference in their stellar mass was compensated by the opposite
SFR difference at late times. Eventually, these two populations of galaxies
have similar stellar mass-to-halo mass ratios.

The cautious reader may have noticed that, in \texttt{EAGLE}, star-forming and
quiescent galaxies have similar halo mass growth history, but end up with
similar descendant stellar mass nonetheless. We found central galaxies in these
massive halos in \texttt{EAGLE} have lower SFR than those in the other two
simulations, and especially so when compared to \texttt{L-GALAXIES}.
Consequently, those star-forming galaxies cannot surpass their quiescent
counterparts too much in terms of stellar mass, even if the latter is quenched
at some stage. Therefore, to reproduce observational results, another required
condition is that star-forming galaxies must grow their stellar mass
significantly after cosmic noon.

In addition, we found that a clear dependence of galaxy quenching time on the
descendant halo mass in \texttt{L-GALAXIES}, where central galaxies end up in
massive halos are quenched earlier than those that end up in less massive halos.
This is a manifestation of galaxy downsizing
\citep[e.g.]{cowieNewInsightGalaxy1996,
fontanotManyManifestationsDownsizing2009}, which is also present in
\texttt{TNG}, but absent in the remaining two hydrodynamical galaxy formation
simulations.

\subsection{Stellar conversion efficiency}%
\label{sub:stellar_conversion_efficiency}

We have seen that \texttt{L-GALAXIES} gives distinct stellar mass growth
history for star-forming and quiescent galaxies, compared with three
hydrodynamical galaxy formation models, and it explains that
\texttt{L-GALAXIES} can reproduce observational SHMRs for different populations
of galaxies while others fail to do so.  Another way to understand this result
is through the efficiency of converting baryons into stars, as
Fig.~\ref{fig:paper_conversion_eff} presents the stellar mass-to-halo mass
ratio for star-forming and quiescent central galaxies as a function of halo
mass for four galaxy formation models. Despite the resemblance of their overall
shape, which peaks around $10^{12}h^{-1}M_\odot$ and is suppressed at both
low-mass and high-mass ends, star-forming galaxies have higher ratios than
quiescent galaxies by $\approx 0.3$ dex in \texttt{L-GALAXIES}, while all three
hydrodynamical galaxy formation simulations give almost identical ratios for
these two populations of galaxies. In other words, star-forming galaxies are
much more efficient in converting baryons into stars than quiescent galaxies in
\texttt{L-GALAXIES}, while, in the other three models, the stellar conversion
efficiencies are close to each other for these two populations of galaxies.
Apparently, the formation of galaxies in these hydrodynamical simulations is
self-regulated. \citep{boothCosmologicalSimulationsGrowth2009,
schayePhysicsDrivingCosmic2010, schayeEAGLEProjectSimulating2015}, so that
these two populations of galaxies end up with similar stellar mass conversion
efficiency.

\section{Discussion}%
\label{sec:discussion}

\subsection{How to relate star formation history to halo growth history?}%
\label{sub:how_to_relate_star_formation_history_to_halo_growth_history_}

How to relate the growth of galaxies to their host dark matter halos plays an
essential role in modeling galaxy formation and evolution. In fact, many
empirical models attempt to directly link these two types of histories to
populate galaxies into dark matter halos/subhalos
\citep[e.g.][]{conroyCONNECTINGGALAXIESHALOS2009,
    behrooziCOMPREHENSIVEANALYSISUNCERTAINTIES2010, luEmpiricalModelStar2014,
moTwophaseModelGalaxy2024}. Here we want to show that the observed SHMRs for
star-forming and quiescent galaxies can put a stringent constraint on these
models. As illustrated in Fig.~\ref{fig:figures/paper_illustration2}, there are
three different scenarios. In \texttt{Scenario-I}, quiescent galaxies live in
early-formed halos \citep[see][]{hearinDarkSideGalaxy2013,
hearinDarkSideGalaxy2014, wangRelatingGalaxiesDifferent2023}, which is
exemplified by three hydrodynamical galaxy formation simulations in this work.
At high redshift, the progenitors of quiescent galaxies grow faster since these
galaxies are star-forming at early times and they have more available baryons
due to more massive halos they reside in. Later on, once these galaxies are
quenched at some point, their stellar mass growth will be suppressed, letting
their star-forming counterparts catch up. Finally, these two populations of
galaxies will end up with similar stellar mass when evolved to $z=0$, as we
have seen in Fig.~\ref{fig:paper_evolution_history} for the three
hydrodynamical galaxy formation simulations.

\texttt{Scenario-II} assumes that the stellar growth history only weakly
relates to the halo assembly history
\citep[e.g.][]{pengMassEnvironmentDrivers2010, pengMASSENVIRONMENTDRIVERS2012,
manFundamentalRelationHalo2019, lyuHalosGalaxiesVII2023, lyu2024}, which is
exemplified by \texttt{L-GALAXIES}. In this case, the progenitors of both
star-forming and quiescent galaxies grow in a similar pace at high redshift
since they are all star-forming and live in similar dark matter halos. Then,
the progenitors of quiescent galaxies must be quenched prior to $z=0$ and, once
that happens, their stellar mass growth will be suppressed and result in lower
stellar mass than their star-forming counterparts, just as we have seen in
Fig.~\ref{fig:paper_evolution_history} for \texttt{L-GALAXIES}. This scenario
could produce the stellar mass difference between star-forming and quiescent
central galaxies at fixed halo mass required by observations.

The last one, \texttt{Scenario-III}, assumes that quiescent galaxies live in
late-formed halos \citep[e.g.][]{cuiOriginGalaxyColour2021,
wangLateformedHaloesPrefer2023, moTwophaseModelGalaxy2024}. In this case, the
stellar mass of star-forming central galaxies is larger than their quiescent
counterparts to a higher degree. This is because not only the stellar mass
growth of quiescent galaxies was suppressed later on, but also the progenitors
of quiescent galaxies have lower stellar mass than those star-forming
counterparts at high redshift due to less massive progenitor halo mass and less
available baryons.

Qualitatively speaking, observational results presented in \S\,\ref{sec:data}
support the last two scenarios, while the scenario in which quiescent central
galaxies that live in early-formed halos are strongly disfavored. Nevertheless,
we notice that if those quiescent galaxies in early-formed halos are quenched
at very early time, they can still end up with lower stellar mass than those
star-forming galaxies in late-formed halos, as shown the pink dashed line in
the bottom-left panel of Fig.~\ref{fig:figures/paper_illustration2}. This
scenario resembles the evolutionary tracks of massive galaxies in the
\texttt{TNG} simulation shown in Fig.~\ref{fig:paper_evolution_history}.
Nonetheless, the stellar mass difference between star-forming and quiescent
central galaxies generated is only $\lesssim 0.1 \rm dex$, which is much lower
than that of the other two scenarios, e.g. $\gtrsim 0.3 \rm dex$ for
\texttt{L-GALAXIES} in Fig.~\ref{fig:paper_evolution_history}, and further
model adjustment is required the match the observational results.  Since the
stellar mass difference at $z=0$ is determined by both the quenching time and
how central galaxy quenching correlates to the halo assembly history, further
discrimination among these scenarios requires not only an accurate measurement
of stellar mass difference between quiescent and star-forming galaxies at fixed
halo mass, but also the measurement of central galaxy quenching and its
dependence on stellar mass and halo mass at high redshift.

\section{Summary}%
\label{sec:summary}

The stellar mass-halo mass relation for central galaxies, as one of the most
fundamental scaling relations in galaxy formation and evolution, has not only
put stringent constraints on galaxy formation models, but also elicited
numerous observational and theoretical work on understanding the feedback
processes that regulate galaxy evolution. Regarding these impactful findings,
measurements of SHMRs for different populations of galaxies have been initiated
and refined in the last decade. Despite the innate systematics of different
methods, all these studies drew on the same conclusion: at fixed stellar mass,
red and quiescent central galaxies tend to live in more massive halos than blue
and star-forming central galaxies at $M_*\gtrsim 10^{10}h^{-1}\rm M_\odot$.
However, the scientific implication of this result is overlooked and
inadequately discussed. In this work, we intend to initiate the discussion on
testing galaxy formation models using the SHMRs for different populations of
central galaxies. Our main results are summarized as follows:

\begin{enumerate}

    \item We compare observational results, in which red/quiescent central
        galaxies prefer to occupy more massive halos than blue/star-forming
        central galaxies at fixed stellar mass, with four state-of-the-art
        galaxy formation models, including one semi-analytical model
        (\texttt{L-GALAXIES}) and three hydrodynamical simulations
        (\texttt{TNG}, \texttt{Illustris}, and \texttt{EAGLE}). We find that
        only \texttt{L-GALAXIES} is consistent with observational results,
        while all three hydrodynamical simulations produce almost identical
        SHMRs for star-forming and quiescent galaxies (see
        Fig.~\ref{fig:paper_compare}). These observational results also imply
        that blue and star-forming central galaxies have higher stellar mass
        than that of red and quiescent central galaxies at fixed halo mass.

    \item We inspect the star formation history and halo growth history for
        halos hosting star-forming and quiescent galaxies, and find that the
        relationship between these two histories is very different for
        \texttt{L-GALAXIES} and the other three hydrodynamical simulations. In
        \texttt{L-GALAXIES}, star formation histories are weakly correlated to
        halo growth histories, and it can successfully reproduce observational
        results. However, in three hydrodynamical simulations, star formation
        histories are strongly correlated to halo assembly histories and
        central galaxies in early-formed halos are more likely to be quenched
        when evolving to $z=0$, resulting in similar stellar mass for star-forming
        and quiescent galaxies at $z=0$, which is against observational results
        (see Figs.~\ref{fig:paper_evolution_history},
        \ref{fig:paper_conversion_eff}, and \ref{fig:figures/paper_illustration2}).

    \item We conclude that the observed SHMRs for star-forming and quiescent
        galaxies support galaxy formation models in which quenching only weakly
        correlates with halo assembly histories, and in which the stellar mass
        of star-forming galaxies can increase significantly since cosmic noon
        ({\tt Scenario-II} in Fig.~\ref{fig:figures/paper_illustration2}),
        while models in which quenching prefer to happen in early-formed halos
        ({\tt Scenario-I} in Fig.~\ref{fig:figures/paper_illustration2}) are
        not very favored given the results in this work. Nevertheless, it is
        still possible for this scenario to produce SHMR that fits the
        observational result by making galaxies in early-formed halos quenched
        earlier, and further model adjustment is required to achieve that.

\end{enumerate}

This work focuses on a qualitative comparison of observational results and galaxy
formation models, and it already puts stringent constraints on galaxy formation
models. To fully release the constraining power of SHMRs for different
populations of galaxies, more accurate observational measurements based on
larger samples is demanding, especially the measurements for individual systems.

\section*{Acknowledgements}

The authors thank the anonymous referee for their helpful comments that
improved the quality of the manuscript. The authors thank Yangyao Chen, Houjun
Mo, and Fangzhou Jiang for their helpful comments and constructive suggestions.
The authors acknowledge the Tsinghua Astrophysics High-Performance Computing
platform at Tsinghua University for providing computational and data storage
resources that have contributed to the research results reported within this
paper. This work is supported by the National Science Foundation of China
(NSFC) Grant No. 12125301, 12192220 and 12192222, and support from the New
Cornerstone Science Foundation through the XPLORER PRIZE.

\section*{Data availability}

The data underlying this article will be shared on reasonable request to the
corresponding author. The computation in this work is supported by the HPC
toolkit \specialname[HIPP] \citep{hipp}.

%%%%%%%%%%%%%%%%%%%%%%%%%%%%%%%%%%%%%%%%%%%%%%%%%%
\bibliographystyle{aasjournal}
\bibliography{bibtex.bib}

% \appendix

\end{document}